\documentclass[aps,prl,twocolumn,showpacs,superscriptaddress,10pt]{revtex4-1}

\usepackage{graphicx}
\usepackage{amsmath,amssymb,amstext,amsthm}
\usepackage{bbold}
\usepackage{cancel}
\usepackage{color}
\usepackage{verbatim}
\usepackage{booktabs}
\usepackage[caption=false]{subfig}
\usepackage[english]{babel}

\newcommand{\Z}{\mathbb{Z}}

\newcommand{\id}{\mathbb{1}}

\usepackage[utf8x]{inputenc}

\begin{document}

\title{Novel topological insulators from crystalline symmetries}

\author{Alexander Lau}
\affiliation{Kavli Institute of Nanoscience, Delft University of Technology, P.O. Box 4056, 2600 GA Delft, Netherlands}

\author{Carmine Ortix}
\affiliation{Institute for Theoretical Physics, Center for Extreme Matter and
Emergent Phenomena, Utrecht University, Princetonplein 5, 3584 CC Utrecht,
Netherlands}
\affiliation{Dipartimento di Fisica ``E. R. Caianiello", Universit\`a di Salerno, IT-84084 Fisciano, Italy}

\date{\today}

\begin{abstract} 
We discuss recent advances in the study of topological insulators protected by spatial symmetries by reviewing three representative, theoretical examples. In three dimensions, these states of matter are generally characterized by the presence of gapless boundary states at surfaces that respect the protecting spatial symmetry. We discuss the appearance of these topological states both in crystals with negligible spin-orbit coupling and a fourfold rotational symmetry, as well as in mirror-symmetric crystals with sizable spin-orbit interaction characterized by the so-called \emph{mirror Chern number}. Finally, we also discuss similar topological crystalline states in one-dimensional insulators, such as nanowires or atomic chains, with mirror symmetry. There, the prime physical consequence of the non-trivial topology is the presence of quantized end charges.
\end{abstract}

\maketitle

The study of topological phases of matter has brought to light a myriad of exceptional features and remarkable effects which have not only broadened our 
fundamental
knowledge of 
quantum states of
matter in general but could also lead to a whole new range of technologies and applications. 
Generally speaking, topological
states of matter are novel 
quantum
phases that elude the celebrated Landau theory of phase transitions and can be described using the mathematical language of topology. A topological phase is characterized by a nonzero topological invariant which is, in contrast to order parameters of conventional phases, a global quantity and assumes only discrete, quantized values. This leads, in turn, to a characteristic quantization of physical observables and to the presence of topologically protected surface or boundary states.

Depending on the dimensionality and the symmetries of the systems under consideration, different topological classes and invariants are possible. This insight has
led to the celebrated Altland-Zirnbauer classification of topological insulators and superconductors
~\cite{Zir96,AlZ97,HHZ05,SRF08,Kit09,SRF09,RSF10,Lud15}, which, however,
takes into account only non-spatial symmetries: time-reversal, particle-hole and chiral symmetry. The topological phases occurring in the Altland-Zirnbauer table are commonly referred to as ``strong'' since their protecting symmetries are \emph{intrinsic} properties of the systems and cannot be represented by unitary operators that commute with the Hamiltonian. As a consequence, topologically protected states appear on \emph{all} surfaces of the system. Furthermore, these states are also robust against
weak disorder because those are spatial modifications of the system and, thus, do not affect a non-spatial symmetry.

In the search for novel topological phases of matter, the notion of topological protection was relaxed to also include ordinary spatial symmetries represented by unitary operators. This led to the discovery of \emph{topological crystalline insulators} (TCI)~\cite{Fu11,AnF15}: novel states of matter whose topological nature arises from crystal symmetries. Those symmetries are point-group symmetries, such as inversion~\cite{HPB11,LuL14}, mirror~\cite{HLL12,CYR13,AFG14,LBO16} and rotation~\cite{Fu11}; space-group symmetries~\cite{WAC16,LZV14,SSG15,SSG16}, such as glide planes and screw axes; or a combination of them~\cite{SMJ12,KBW17,PVW17,WBW18}. Recently, this concept has even been extended to include magnetic space groups~\cite{FGB14,ZhL15,WPV18}. 
The discovery of TCIs has opened the door to a plethora of new topological phases based on the richness and complexity of crystal structures. Topological crystalline insulators are ``weaker'' than their strong relatives. The reason is twofold. First, crystal symmetries are susceptible to disorder. Therefore, topological features are expected to persist only if the protecting crystal symmetry is preserved on average~\cite{ShS14,LBO16}. Second, \emph{not all} surfaces of a TCI can accommodate topological surface states, but only those that do not break the protecting crystal symmetry. 
To prepare and find novel TCI phases, it is useful to extend the standard Altland-Zirnbauer table to include also crystalline symmetries. This has been done for several cases, such as systems with inversion~\cite{LuL14} or reflection symmetry~\cite{CYR13,ShS14}. In general, the resulting tables are much more involved than the standard Altland-Zirnbauer table due to the various possible relations between the non-spatial symmetry operators and the additional unitary symmetries. The corresponding invariants are typically derived from strong topological invariants associated with symmetry-invariant Brillouin zone (BZ) cuts as we will see below. However, later on we will also see that this is not always the case.

In this short review, we are going to discuss three selected examples of topological crystalline insulators. By means of these pedagogical examples we illustrate how crystalline symmetries can lead to novel topological phases, how new topological invariants can be defined, and what features can arise as a consequence. For a broader and more in-depth recapitulation of the field, we refer the reader to the review by Ando and Fu~\cite{AnF15}.

\section{Topological crystalline insulators with rotational symmetry}

\begin{figure}[t]\centering
\includegraphics[width=1.0\columnwidth]
{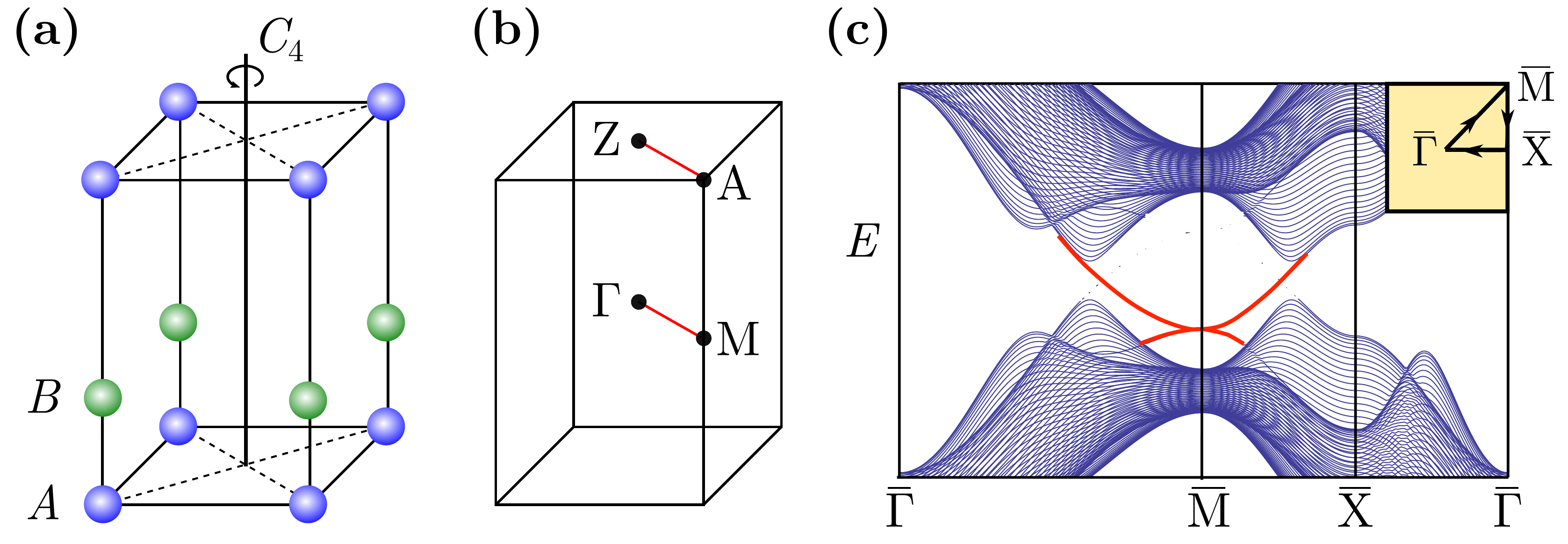}
\caption{(a) Bipartite tetragonal lattice with a $C_4$ axis. (b) Corresponding bulk BZ with the four bulk momenta invariant under $C_4$ rotation. The red lines indicate possible paths for the calculation of the topological invariant $\nu_0$. (c) Band structure along high-symmetry lines in the (001) surface BZ as presented in Ref.~\cite{Fu11}. A quadratic surface band is highlighted in red.}
\label{fig:TCI_with_C4}
\end{figure}

The notion of ``topological crystalline insulators'' was first introduced by Fu using the example of a system with $C_4$ rotational symmetry~\cite{Fu11}. To obtain a minimal model for a TCI, he considered a system of spinless electrons on a tetragonal lattice with a unit cell consisting of two inequivalent atoms $A$ and $B$ stacked along the $c$ axis [see Fig.~\ref{fig:TCI_with_C4}(a)]. For this system, Fu derived a general four-band tight-binding model taking into account only $p_x$ and $p_y$ orbitals of the electrons. Due to the symmetry of the lattice and the chosen orbitals, the model has a natural $C_4$ symmetry with respect to a rotation about the $z$ axis. Furthermore, the model has time-reversal symmetry with $\Theta=K$ corresponding to complex conjugation.

This model features a gapped phase in a finite parameter range. Most remarkably, the (001) surface, which preserves the $C_4$ symmetry, exhibits surface states traversing the entire bulk energy gap in this phase as shown in Fig.~\ref{fig:TCI_with_C4}(c). Moreover, the surface states are doubly degenerate at the $\bar{M}$ point of the surface BZ. This degeneracy is not accidental. The $\bar{M}$ point is a fixed point under fourfold rotation and the two degenerate states form a two-dimensional (2D) irreducible real representation of $C_4$. Hence, the degeneracy is protected by symmetry and cannot be removed. Note that there is no Kramers theorem to enforce the degeneracy because $\Theta^2=+1$.

Close to the $\bar{M}$ point, the two bands of surface states can be represented in terms of $p_x$ and $p_y$ orbitals. In this representation, $C_4$ rotation can be represented by $e^{i\sigma_y \pi/4}$, where $\sigma_y$ is a Pauli matrix. In the presence of $C_4$ and $\Theta$ symmetry, it can be shown that, to leading order, the Hamiltonian of the surface states must be of the form
\begin{equation}
H(k_x,k_y) = \frac{k^2}{2m_0}\,\id + \frac{k_x^2 - k_y^2}{2m_1}\,\sigma_z + \frac{k_x k_y}{2m_2}\,\sigma_x.
\end{equation}
Hence, the symmetry constrains suppress the linear order and lead to a \emph{quadratic} dispersion around the $\bar{M}$ point, as opposed to a linear Dirac dispersion for ``conventional'' topological insulators.

Moreover, these surface states are topologically protected. This can be seen as follows: due to the symmetry of the model, (001) surface states at the $C_4$-invariant momenta $\bar{M}$ and $\bar{\Gamma}$ must be doubly degenerate. Hence, in analogy with quantum spin Hall insulators (QSHIs) in 2D, there are two topologically distinct ways of connecting these doublets with each other and with the valance and conduction bands. This gives rise to a $\Z_2$ classification, where in the nontrivial phase the surface bands cross the Fermi level an \emph{odd} number of times along a path connecting $\bar{\Gamma}$ and $\bar{M}$. Note that other surfaces do not feature topological surface states because they break $C_4$ symmetry.

From a more general perspective, the topological nature of a spinless, time-reversal invariant (TRI) insulator with $C_4$ symmetry can be understood in the following way. There are four $C_4$-invariant momenta $\mathbf{k}_i$ in the three-dimensional (3D) BZ of the system, namely $\Gamma=(0,0,0)$, $M=(\pi,\pi,0)$, $A=(\pi,\pi,\pi)$, and $Z=(0,0,\pi)$ [see Fig.~\ref{fig:TCI_with_C4}(b)]. At these points, the Bloch Hamiltonian $H(\mathbf{k}_i)$ of the system commutes with the unitary operator $U$ representing a $C_4$ rotation. Hence, the energy states at $\mathbf{k}_i$ can be chosen to be eigenstates of fourfold rotation with possible rotation eigenvalues $1$, $-1$, $i$, and $-i$. In addition, the momenta $\mathbf{k}_i$ are invariant under time-reversal. Due to time-reversal symmetry with $\Theta=K$ and $\Theta^2=+1$, this imposes a reality condition. More specifically, the Bloch Hamiltonian $H(\mathbf{k}_i)$ at these points must be real and its eigenstates can always be chosen to be real. Under these conditions, group theory tells us that whenever there is a state with rotation eigenvalue $\pm i$, there must also be another, degenerate state with eigenvalue $\mp i$. The reason is that the cyclic group $C_4$ is a rather peculiar group. It has four one-dimensional (1D) irreducible representations. However, the two representations with $C_4$ eigenvalues $\pm i$, which are commonly grouped under the representation label $E$, together form a so-called \emph{separable degenerate} representation~\cite{book:Ket89}. It can be shown that these representations always appear together and give rise to a degenerate doublet of states.
Furthermore, such a doublet gives rise to an effective Kramers theorem with respect to the operator $\tilde{\Theta}\equiv U\Theta$, because  $\tilde{\Theta}^2 = U\Theta U\Theta = U^2 \Theta^2 = -1$,
where we have used that, with eigenvalues $\pm i$, the rotation operator $U$ squares to $-1$. Note that we do not have an effective Kramers theorem for bands with rotation eigenvalues $\pm 1$, because the corresponding rotation operator would square to $+1$.

For simplicity, let us assume that all occupied bands transform as doublets under fourfold rotation. In this case, we can follow the steps of the derivation of the Fu-Kane invariant for QSHIs in 2D~\cite{FuK06}, just with the operator $\Theta$ replaced by $\tilde{\Theta}$ and with paths in $\mathbf{k}$ space connecting the $C_4$-invariant momenta $\mathbf{k}_i$. Finally this leads to $\Z_2$ topological invariants $\nu_0$ and $\nu_{\Gamma M}$, $\nu_{AZ}$ of the following form~\cite{Fu11}:
\begin{eqnarray}
(-1)^{\nu_0} &=& (-1)^{\nu_{\Gamma M}} (-1)^{\nu_{AZ}},\label{eq:C4_index_relation}\\
(-1)^{\nu_{\mathbf{k}_1\mathbf{k}_2}} &=&
\exp \bigg(i\int_{\mathbf{k}_1}^{\mathbf{k}_2} d\mathbf{k}\cdot \mathbf{A}(\mathbf{k})\bigg)
\frac{\mathrm{Pf}[w(\mathbf{k}_2)]}{\mathrm{Pf}[w(\mathbf{k}_1)]},
\label{eq:C4_index_definition}
\end{eqnarray}
where $\mathbf{A}(\mathbf{k})=-i\sum_j \langle u_{j\mathbf{k}}|\partial_\mathbf{k}|u_{j\mathbf{k}}\rangle$
is the $U(1)$ Berry connection and the matrix elements of $w(\mathbf{k}_i)$ are defined as 
$w_{mn}(\mathbf{k}_i) =\langle u_{m\mathbf{k}_i}|U\Theta|u_{n\mathbf{k}_i}\rangle$. This matrix is antisymmetric because $[H(\mathbf{k}_i),U\Theta]=0$ and $(U\Theta)^2=-1$. The line integrals for 
$\nu_{\Gamma M}$ and $\nu_{AZ}$ are along arbitrary paths connecting 
$\Gamma$ with $M$ and $A$ with $Z$,
respectively, that lie within the plane $k_z=0$ and $k_z=\pi$, respectively, as illustrated in Fig.~\ref{fig:TCI_with_C4}(b). It can be shown that, in contrast to QSHIs, $(-1)^{\nu_{\mathbf{k}_1\mathbf{k}_2}}$ is already a gauge-invariant $\Z_2$ quantity, i.e., $\nu_{\Gamma M}$ and $\nu_{AZ}$ define topological invariants for the planes $k_z=0$ and $k_z=\pi$, respectively. Due to the relation in Eq.~\eqref{eq:C4_index_relation} between the three $\Z_2$ invariants, a 3D crystalline topological insulator with time-reversal and $C_4$ symmetry is fully characterized by the strong index $\nu_0$ and one of the weak indices $\nu_{\Gamma M}$ or $\nu_{AZ}$. This is similar to the relation between strong and weak indices for 3D topological insulators with time-reversal symmetry. However, only the strong phase with $\nu_0=1$ gives rise to topological surface states on the (001) surface.

The $\Z_2$ invariants are only defined for doublet bands. Real materials usually have both doublet and singlet bands. Nonetheless, the $\Z_2$ invariants remain well-defined as long as the doublet bands can be energetically separated from the singlet bands, which is usually the case. This shows an important fundamental difference between ``conventional'' topological insulators and topological crystalline insulators. In the latter, there is an interplay between symmetry representations and the topology of the corresponding energy bands. Specifically, in our example the nontrivial topology arises from doublet bands alone whereas singlet bands are always trivial. 

\section{Mirror Chern number for systems with reflection symmetry}
\label{sec:mirror_Chern_number}

\begin{figure}[t]\centering
\includegraphics[width=1.0\columnwidth]
{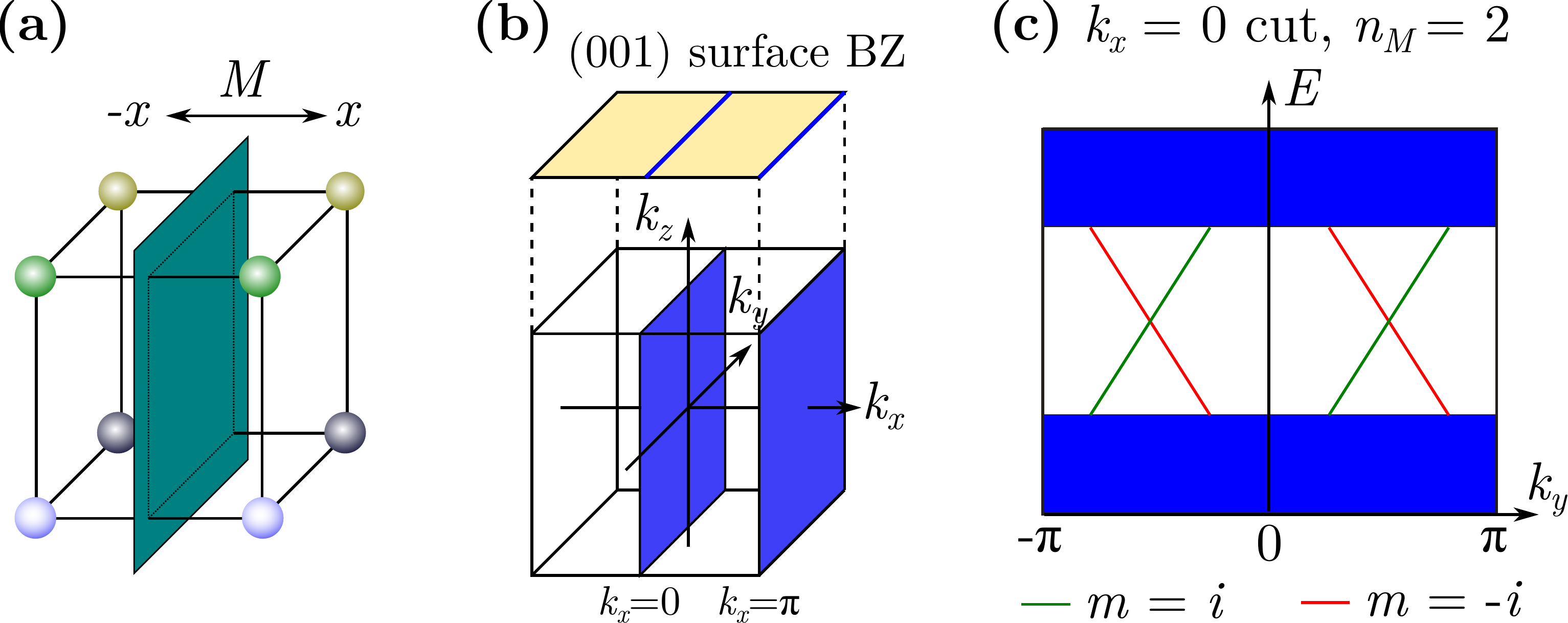}
\caption{(a) Simple cubic lattice with a mirror plane highlighted in turquoise. (b) Corresponding bulk BZ with the two mirror-invariant planes highlighted in blue. The (001) surface BZ is shown on top including the projections of the mirror-invariant planes. (c) Sketch of a nontrivial surface band structure along the projection of the $k_x=0$ mirror-invariant plane.}
\label{fig:TCI_with_mirror}
\end{figure}

Let us consider a 3D insulating crystal with time-reversal symmetry and reflection symmetry with respect to a mirror plane. Without loss of generality, let the mirror plane be parallel to the $yz$ plane, i.e., the mirror operation takes $x$ to $-x$. For simplicity, we further assume that the crystal is a simple cubic lattice [see Fig.~\ref{fig:TCI_with_mirror}(a)]. Due to reflection symmetry, the Bloch Hamiltonian $H(\mathbf{k})$ of the system must satisfy
\begin{equation}
M H(k_x,k_y,k_z)M^{-1} = H(-k_x,k_y,k_z),
\end{equation}
where $M$ is a unitary operator representing the mirror operation. From this equation we immediately see that the Bloch Hamiltonian commutes with the mirror operator $M$ in the mirror-invariant planes $k_x=0$ and $k_x=\pi$ [see Fig.~\ref{fig:TCI_with_mirror}(b)]. Therefore, in these planes the Bloch Hamiltonian decomposes into two blocks corresponding to the mirror eigenvalues $m=+i$ and $m=-i$, respectively. Since the system has an energy gap, we can associate a Chern number $n_{\pm i}(k_x)$ with each of the blocks, where $k_x=0$ or $\pi$. Moreover, time-reversal symmetry requires $n_{+i} + n_{-i} = 0$. Nevertheless, in analogy with QSHIs, the difference of the two numbers defines a toplogical invariant for each of the mirror-invariant planes. This is the so-called \emph{mirror Chern number}~\cite{TFK08},
\begin{equation}
n_M(k_x) = \frac{1}{2}[n_{+i}(k_x) - n_{-i}(k_x)].
\end{equation}

What are the implications for surface states? First of all, we already know that surface states can only be protected on surfaces that do not break reflection symmetry. In our example, such a surface is for instance the (001) surface. In the corresponding surface BZ the lines with $k_x=0$ and $k_x=\pi$ correspond to surface projections of the mirror-invariant planes above [see Fig.~\ref{fig:TCI_with_mirror}(b)]. Let us assume that $n_M$ is nontrivial for the $k_x=0$ plane. Thus, the nontrivial blocks of the Bloch Hamiltonian in this plane give rise to $n_M$ pairs of counterpropagating surface states along $k_x=0$ by bulk-boundary correspondence [see Fig.~\ref{fig:TCI_with_mirror}(c)]. At the crossing point between two counterpropagating states, it is not possible to couple the states in order to open an energy gap since they belong to two different reflection-symmetry sectors. Hence, the crossing point is protected by symmetry. Away from $k_x=0$, the states no longer have a well-defined mirror eigenvalue and the degeneracy is lifted. This gives rise to $n_M$ topologically protected surface Dirac cones. In contrast to 3D TRI topological insulators, where the Dirac cones are pinned to TRI momenta, the surface Dirac cones of a system with nontrivial mirror Chern numbers can in principle be anywhere along the line $k_x=0$ while respecting time-reversal symmetry~\cite{HLL12}.

\begin{figure}[t]\centering
\includegraphics[width=1.0\columnwidth]
{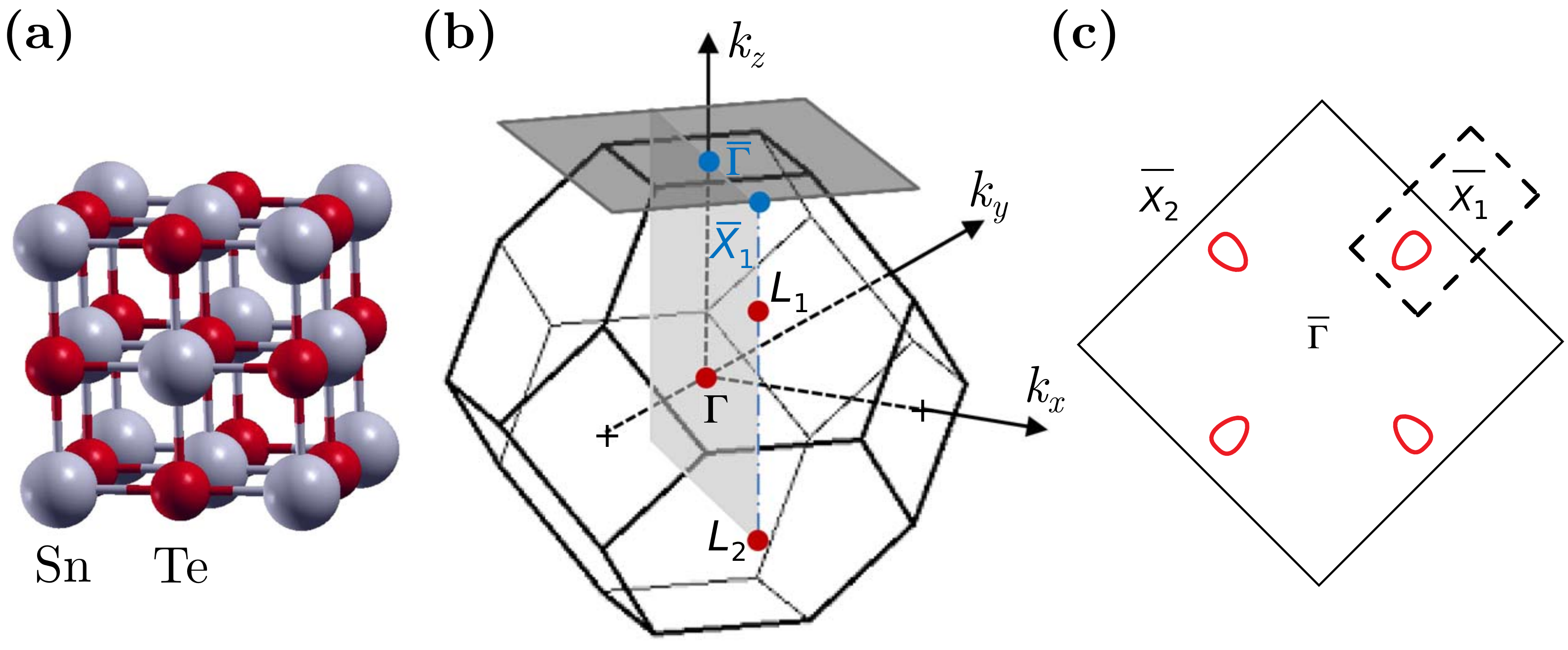}
\caption{Lattice and BZs of the topological crystalline insulator SnTe as presented in Ref.~\cite{HLL12}: (a) face-centered cubic lattice. (b) Corresponding BZ with high-symmetry points. Also indicated is the (001) surface BZ and one of the mirror-invariant planes. (c) Fermi surface of the (001) surface with four topologically protected Dirac-cone pockets.}
\label{fig:TCI_with_mirror_SnTe}
\end{figure}

A material realization of a topological crystalline insulator with nonzero mirror Chern number is SnTe~\cite{HLL12,TRS12}. The crystal structure of SnTe is face-centered cubic (rocksalt), as illustrated in Fig.~\ref{fig:TCI_with_mirror_SnTe}(a). Moreover, the material is an insulator with trivial $\Z_2$ invariants. However, the (011) mirror plane, and other planes equivalent by symmetry, give rise to a nontrivial mirror Chern number of $n_M=-2$ and, thus, to topologically protected Dirac cones on certain surfaces~\cite{HLL12}. Remarkably, the \emph{total} number of protected Dirac cones depends on the considered surface, which is in stark contrast to ``conventional'' topological insulators in 3D. 

The (001) surface is symmetric about the two equivalent (011) and (0-11) mirror planes. In the surface BZ, these planes are projected onto two perpendicular lines going through the $\bar{X}_1$ and $\bar{X}_2$ points. Since we have mirror Chern numbers of $n_M=-2$ associated with \emph{both} lines, there are in total $2\times 2=4$ surface Dirac cones around the $\bar{\Gamma}$ point of the surface BZ [see Fig.~\ref{fig:TCI_with_mirror_SnTe}(c)].
Similarly, we can deduce the number of surface Dirac cones for other terminations. The (111) surface preserves three equivalent mirror planes which are projected onto three mirror-invariant lines in the corresponding surface BZ. Hence, there are $2\times 3=6$ topologically protected surface Dirac cones on the (111) surface of SnTe. Finally, the (110) surface is symmetric about one mirror plane giving rise to two surface Dirac cones.

\section{One-dimensional topological insulators with time-reversal symmetry}

As we have seen in the previous sections, crystal symmetries can lead to entirely novel topological phases of matter.  
In particular, classes of systems that are trivial according to the standard Altland-Zirnbauer classification of topological insulators might reveal a nontrivial topological nature once crystal symmetries are imposed.

In this light, we are going to turn our attention to one of the most-studied classes of the Altland-Zirnbauer table, namely the symplectic class AII~\cite{RSF10}. This class contains all time-reversal symmetric systems whose time-reversal operator $\Theta$ squares to $-1$, and thus concerns systems of spin-$1/2$ electrons with time-reversal symmetry. Most importantly, the famous QSHIs in 2D and the $\Z_2$ topological insulators in 3D fall into this class. However, in one dimension this class does not allow for a nontrivial topology in the scope of the standard Altland-Zirnbauer classification. Nevertheless, as we will review below, it can be shown~\cite{LBO16} 
that the presence of a crystalline symmetry, such as reflection symmetry, gives rise to a class of one-dimensional TRI crystalline topological insulators beyond the standard Altland-Zirnbauer scheme.

\begin{figure}[t]\centering
\includegraphics[width=1.0\columnwidth]
{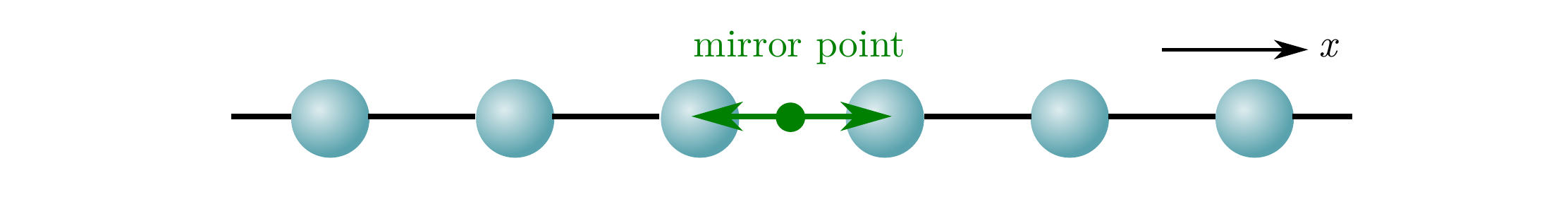}
\caption{One-dimensional lattice with a mirror point.}
\label{fig:1D_chain}
\end{figure}

For that purpose, let us now look into generic systems of fermions with spin one half subject to a 1D crystalline potential. Furthermore, let us impose time-reversal symmetry, as well as reflection symmetry with respect to a 1D mirror point (see Fig.~\ref{fig:1D_chain}). Due to the periodicity of the lattice, such systems can be described by a Bloch Hamiltonian $H(k)$, with the crystal momentum $k\in(-\pi,\pi]$, which then satisfies $\Theta H(k) \Theta^{-1} = H(-k)$, where $\Theta=(\id \otimes i s^y)K$ is the antiunitary time-reversal operator, and $M H(k)M^{-1} = H(-k)$, with the unitary reflection operator $ M={\cal I} \otimes i s^x$. Here, the $s^i$ are Pauli matrices associated with the system's spin degree of freedom, while ${\cal I}$ corresponds to spatial inversion with respect to other degrees of freedom the system might have.
Furthermore, we have that  $\Theta^2=-1$, $M^2=-1$, and $[\Theta,M]=0$. Given these general properties, our system allows for a classification in terms of a $\Z_2$ topological invariant, according to the extended Altland-Zirnbauer table augmented by mirror symmetry~\cite{CYR13,ShS14}. In particular, this invariant can be formulated using the concept of charge polarization~\cite{LBO16}. 

\subsection{Topological invariant for topological mirror insulators in one dimension}

The \emph{total} charge polarization associated with the 1D Bloch Hamiltonian
$H(k)$ can be formulated as~\cite{Zak89,KiV93,Res94,Res00}:
\begin{equation}
P_\rho = \frac{1}{2\pi}\int_{-\pi}^{\pi} dk\: A(k),
\label{eq:polarization}
\end{equation}
with the $U(1)$ Berry connection
\begin{equation}
A(k)=i\sum_n \langle u_{k,n}|\partial_k|u_{k,n}\rangle,
\label{eq:full_berry_connection}
\end{equation}
where  $|u_{k,n}\rangle$ is the lattice-periodic part of a Bloch state at
momentum $k$ and band index $n$, and the sum is over all occupied bands. Note that we have set both the electronic charge and the lattice constant to unity. It can be shown that the quantity above is only well-defined up to an integer. Furthermore, it can generally assume any value and can, hence, not serve as a topological invariant.

To define topological invariants for TRI topological insulators in 2D and 3D, Fu and Kane introduced the notion of \emph{partial polarization}~\cite{FuK06}. Remarkably, also for systems in 1D the partial polarization can be used to define a topological invariant.

In insulators with time-reversal symmetry, due to Kramers' theorem all $2N$ occupied bands can be divided into $N$ pairs connected by time reversal~\cite{FuK06}. Such a bipartition can be established by defining
\begin{equation}
|u_{-k,\alpha}^\mathrm{I}\rangle = -e^{i\chi_\alpha(k)}\, \Theta
|u_{k,\alpha}^{\mathrm{II}}\rangle,
\label{eq:I_II_partition}
\end{equation}
where $\Theta$ is the time-reversal operator from above,
$\alpha=1,\ldots,N$, and I,~II are the two time-reversed channels. With this, the
partial polarizations are simply the polarizations associated with the two
channels, namely,
\begin{equation}
P^s = \frac{1}{2\pi}\int_{-\pi}^{\pi} dk\: A^{s}(k),\:\:
s=\mathrm{I,II},
\label{eq:def_partial_pol}
\end{equation}

We can restrict our consideration to $P^\mathrm{I}$ since the two partial polarizations are connected by $P^\mathrm{I} = P^\mathrm{II} \textrm{ mod } 1$.
Moreover, it can be shown that, given the system preserves mirror symmetry with
$M H(k) M^{-1} = H(-k)$ and $[M,\Theta]=0$,
we also have $P_\mathrm{I} = -P_\mathrm{I}\,
\mathrm{mod}\,1$~\cite{LBO16}. 
Hence, the partial polarization $P^I$ can only be $0$ or $1/2$, up to an integer, and a $\Z_2$ topological invariant is naturally defined by setting $\nu=2 P^{s} \mathrm{mod}\, 2 \equiv 0,1$. As a consequence, there exist two topologically distinct phases: trivial insulators with $\nu=0$, and topological mirror insulators with $\nu=1$. It is not possible to connect these phases by adiabatic transformations that conserve the defining symmetries and that keep the bulk gap open. The corresponding topological invariant can be written explicitly as~\cite{LBO16},
\begin{equation}
\nu := \frac{1}{\pi} \bigg[ \int_0^{\pi} dk\:A(k) + i\,\log\Big(
\frac{\mathrm{Pf}[w(\pi)]}{\mathrm{Pf}[w(0)]} \Big)\bigg]
\,\mathrm{mod}\,2.
\label{eq:invariant}
\end{equation}
The quantity $w_{\mu\nu}(k)= \langle u_{-k,\mu}|\Theta|u_{k,\nu}\rangle$ is a $U(2N)$ matrix, where $2N$ is the number of occupied energy bands. Furthermore, this matrix is antisymmetric at $k=0,\pi$ and can therefore be assigned a Pfaffian $\mathrm{Pf}(w)$.

It is important to emphasize that the topological invariant of
this class of systems cannot be determined from the knowledge of the electronic wavefunctions only at the mirror invariant momenta, as is the case for instance for the mirror Chern insulators discussed above.
In particular, this implicates that topological gap closing and reopening transitions generally occur away from high-symmetry points in the 1D BZ.  

Furthermore, also the bulk-boundary correspondence is rather unconventional in these systems, and
can be understood as follows. In the standard Altland-Zirnbauer table, all nontrivial classes in 1D have either particle-hole or chiral symmetry. Thus, by bulk-boundary correspondence, nontrivial insulators (or superconductors) in these classes feature topological end states bound to zero energy which are protected by one or both of these two symmetries. By allowing for additional symmetries, also systems with neither particle-hole nor chiral symmetry can become topologically nontrivial, as we have seen in this section. As a consequence, topological end states are no longer bound to appear at zero-energy and can even be ``pushed'' out of the bulk-energy gap: the presence of end states is no longer a protected feature.

\begin{figure}[t]\centering
\includegraphics[width=1.0\columnwidth]
{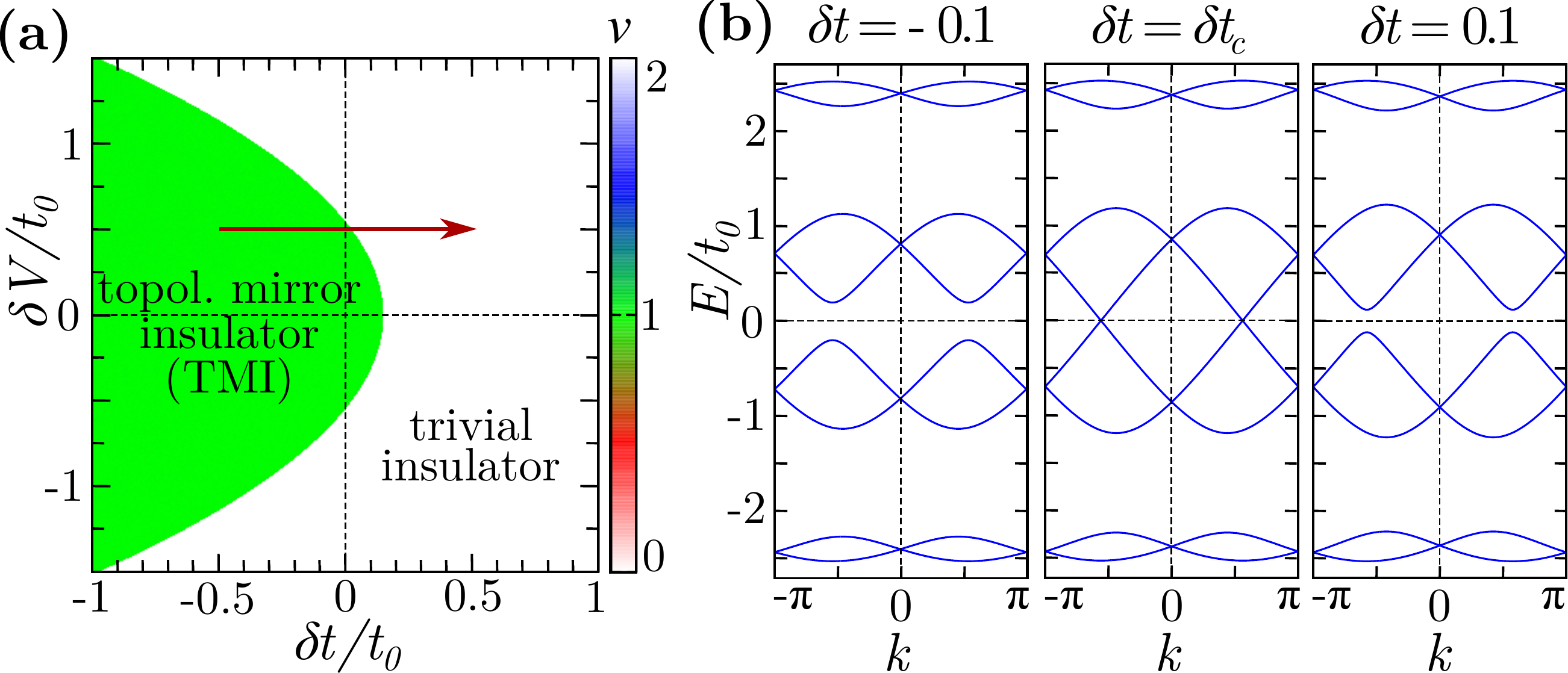}
\caption{Phase diagram and spectra of the spin-orbit coupled, mirror-symmetric 
AAH model with periodic boundary conditions and $\alpha=\gamma=1/2$, $\beta=1/4$, $V_0=0$,
$\lambda_0=0.5t_0$, $\delta\lambda=-0.3t_0$, $\phi_t=\phi_\lambda=\pi$: (a)
half-filling phase diagram of the bulk for $\phi_V=-\pi/4$. The value of the
$\Z_2$ invariant $\nu$ is indicated by pixel color. (b) Bulk band structures for $\delta V=0.5t_0$ and different $\delta t$. The band structures correspond to systems along the red arrow in (a). (Adapted from Ref.~\cite{LBO16})}
\label{fig:AAH_phase_diagram_and_gap_closing}
\end{figure}

In contrast, the bulk-boundary correspondence in the class of systems considered here manifests itself in the presence of quantized end \emph{charges}. 
In fact, it can be shown that, in general, a system's charge polarization is directly linked to the charge accumulated at its boundary~\cite{KiV93}. We have seen that, for our systems, this polarization is composed of two identical contributions $P^\mathrm{I}$ and $P^\mathrm{II}$. Hence, the net bound charge at each of the two end points of the system is~\cite{LBO16},
\begin{equation}
Q_b\,\mathrm{mod}\,2 = 2P_\mathrm{I} =
\nu.
\label{eq:boundary_charge}
\end{equation}
In other words, the total bound charge can be written in terms of the system's $\Z_2$ invariant $\nu$ and is therefore topologically \emph{protected}. More specifically, the characterizing feature of a topological mirror insulator with $\nu=1$ is the presence of an \emph{odd} number of integer-valued electronic end charges at its mirror-symmetric boundaries.

\subsection{Spin-orbit coupled Aubry-Andr\'e-Harper models: a realization of topological mirror insulators}

Having established the general theory, we will now discuss a model that realizes a topological mirror insulator. In particular, let us consider an Aubry-Andr{\'e}-Harper (AAH) model~\cite{Har55,AuA80,GSS13,LOB15_2} with spin-orbit coupling given by the Hamiltonian below~\cite{LBO16},
\begin{eqnarray}
\mathcal{H}&=&\sum_{j,\sigma} [t_0+\delta t\cos(2\pi\alpha j + \phi_t)]\, c_{j+1,\sigma}^\dagger c_{j\sigma}\nonumber\\
&&{}+ \sum_{j,\sigma} [V_0 + \delta V\cos(2\pi\beta j + \phi_V)]\, c_{j\sigma}^\dagger c_{j\sigma} \nonumber\\
&&{}+ i\sum_{j,\sigma,\sigma'} [\lambda_0 + \delta\lambda\cos(2\pi\gamma j + \phi_\lambda)]\, c_{j+1,\sigma}^\dagger s^y_{\sigma\sigma'}c_{j\sigma'}\nonumber\\
&&{} + \mathrm{h.c.}.
\label{eq:AAH_model_general}
\end{eqnarray}
Here, the operators $c_{j\sigma}^\dagger$
($c_{j\sigma}$) create (annihilate) an electron with spin $\sigma$
($\sigma=\uparrow,\downarrow$)
at lattice site $j$, and the $s^i$ are Pauli matrices. 
The model contains
harmonically modulated nearest-neighbor hopping, 
on-site potentials and SOC. The Hamiltonian of Eq.~\eqref{eq:AAH_model_general} has time-reversal
symmetry whereas the model is mirror symmetric only for certain set of parameter configurations~\cite{LBO16}. 
Experimentally, this model can potentially be studied with ultracold Fermi gases in optical lattices \cite{Bloch05,AAL13,CSH12,WYF12,Zhai15,LLC13,CJS11,WKL15} or in a semiconductor nanowire with Rashba SOC~\cite{LBO16}.

Here, we are going to discuss the model in Eq.~\eqref{eq:AAH_model_general} with parameters $\alpha=\gamma=1/2$, $\phi_t=\phi_\lambda=\pi$, $V_0=0$, and $\beta=1/4$, for which the model preserves reflection
symmetry if $\phi_V=-\pi/4$ or $3\pi/4$.
Fig.~\ref{fig:AAH_phase_diagram_and_gap_closing}(a) shows the values of the $\Z_2$ invariant $\nu$ as a function of the parameters $\delta V$ and $\delta t$ for fixed $\phi_V=-\pi/4$ and with periodic boundary conditions. We observe that $\nu$ is indeed quantized and gives rise to two distinct phases: a trivial phase ($\nu=0$) on the right, and a topological mirror insulator phase ($\nu=1$) on the left. Furthermore, we find that there is a bulk energy gap closing-reopening transition across the phase boundary. Notably, the bulk gap closes at BZ points different from the mirror-invariant momenta [see Fig.~\ref{fig:AAH_phase_diagram_and_gap_closing}(b)]. As pointed out in the previous section, this is in stark contrast to other crystalline topological phases.

\begin{figure}[t]\centering
\includegraphics[width=1.0\columnwidth]
{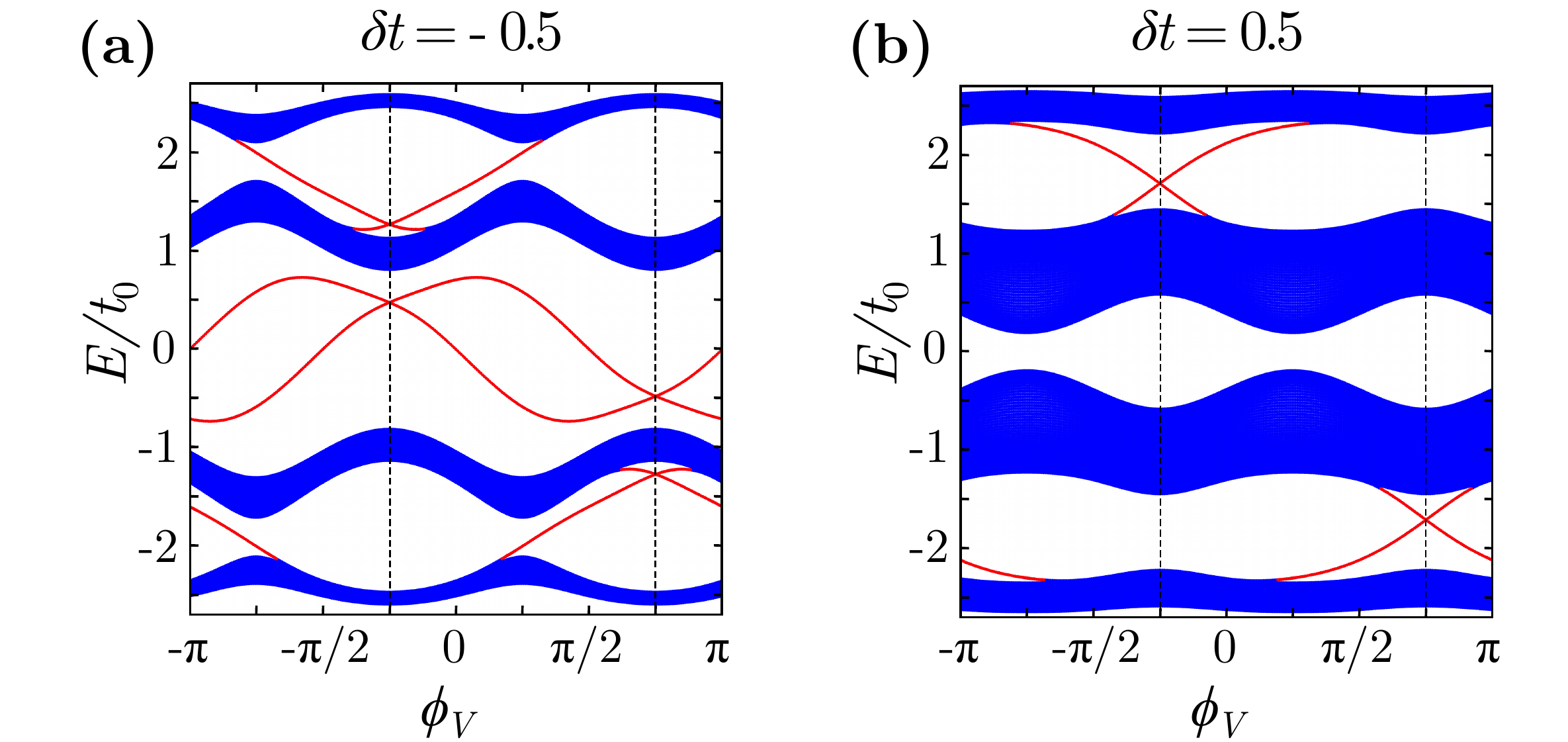}
\caption{Spectra of the spin-orbit coupled 
AAH model with open boundary conditions as a function of $\phi_V$. Other parameters are
$\alpha=\gamma=1/2$, $\beta=1/4$, $V_0=0$, $\delta V=0.5t_0$
$\lambda_0=0.5t_0$, $\delta\lambda=-0.3t_0$, $\phi_t=\phi_\lambda=\pi$: 
(a) $\delta t=-0.5t_0$, (b) $\delta t=0.5t_0$. 
States localized at the ends of the chain are highlighted in red. The values $\phi_V=-\pi/4$ and $3\pi/4$ correspond to a mirror-symmetric chain. (Adapted from Ref.~\cite{LBO16})}
\label{fig:AAH_finite}
\end{figure}

Next, let us look into the bulk-boundary correspondence of the model. For this purpose, let us first discuss its energy spectra for open boundary conditions as a function of $\phi_V$, as shown in Fig.~\ref{fig:AAH_finite}.
As the sweeping parameter is varied, the system passes 
through $\phi_V=-\pi/4$ and $3\pi/4$, at which the model preserves reflection symmetry. Here, the finite, half-filled system features four degenerate in-gap end states as long as the system parameters are in the topological region of Fig.~\ref{fig:AAH_phase_diagram_and_gap_closing}(a). This is shown in Fig.~\ref{fig:AAH_finite}(a). Once mirror symmetry is broken, which is the case away from the $\phi_V$ values specified above, the quadruplet of end states are split into Kramers doublets. We note that the model is time-reversal symmetric for all $\phi_V$. On the contrary, we observe no end states at half filling in the trivial phase [see Fig.~\ref{fig:AAH_finite}(b)].

\begin{figure}[t]\centering
\includegraphics[width=1.0\columnwidth]{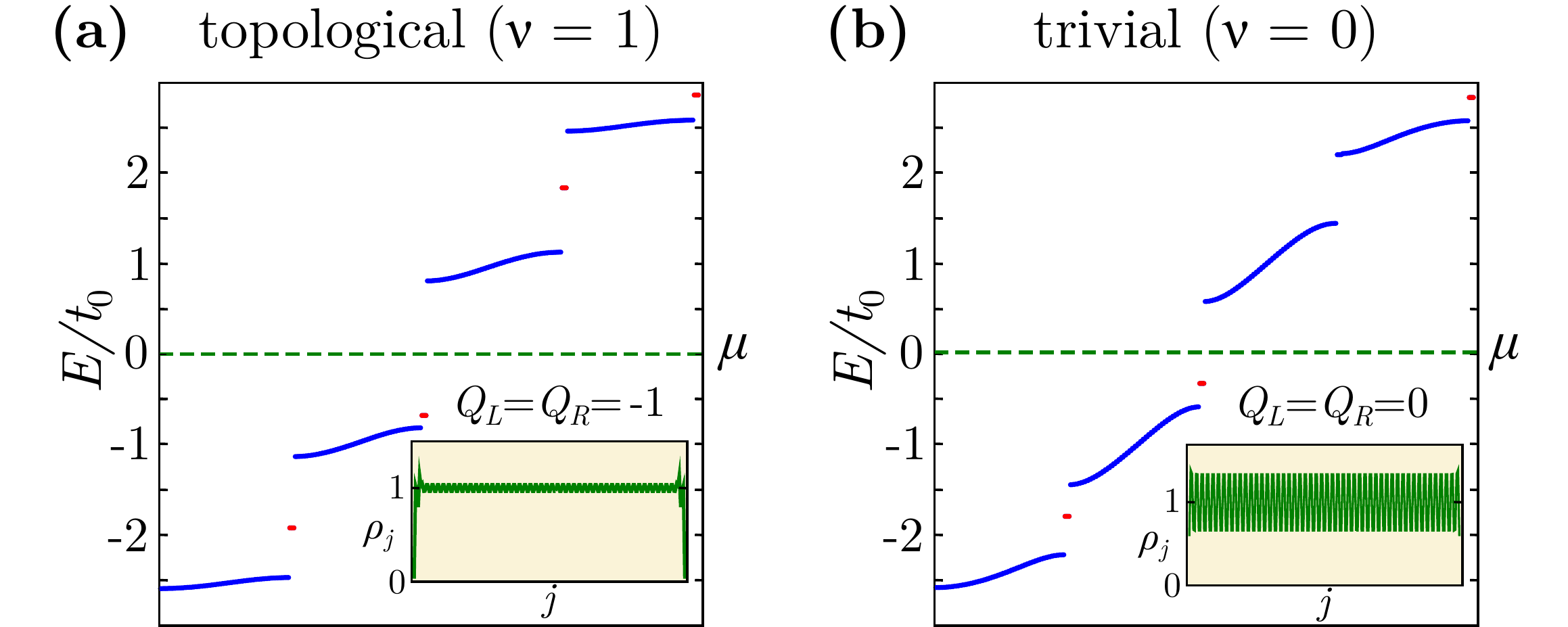}
\caption{Spectra and local charge densities of the
spin-orbit coupled, mirror-symmetric AAH model with open
boundary conditions and $\alpha=\gamma=1/2$, $\beta=1/4$,
$V_0=0$, $\delta V=0.5t_0$, $\lambda_0=0.5t_0$, $\delta\lambda=-0.3t_0$,
$\phi_t=\phi_\lambda=\pi$, $\phi_V=-\pi/4$, and a boundary potential
$V_\mathrm{LR}=0.6t_0$:
(a) topological phase with $\delta
t=-0.5t_0$,
(b) trivial phase with $\delta
t=0.4t_0$.
The main panels show the energy spectra with end states highlighted in red. The dashed line denotes the chemical potential $\mu$ and the corresponding local charge densities $\rho_j$ are presented in
the insets. In addition, the corresponding values of the electronic
end charges $Q_\mathrm{L}$ and $Q_\mathrm{R}$ are displayed. (Adapted from Ref.~\cite{LBO16})}
\label{fig:bands_with_surface_potential}
\end{figure}

However, as discussed in the previous section, the presence of in-gap end states is not a protected feature due to the lack of
chiral symmetry. In particular, this means that 
symmetry-allowed perturbations can move the end
modes out of the bulk energy gap. 
The point above can be demonstrated explicitly by adding a generic on-site boundary
potential $\sum_\sigma V_\mathrm{LR}(c_{1\sigma}^\dagger c_{1\sigma} + c_{L\sigma}^\dagger
c_{L\sigma})$ to the model.
In this case, we observe that the end states in the half-filling gap disappear into the conduction band. At the same time, another quadruplet of end modes moves up from the valence band into the bulk energy gap [see Fig.~\ref{fig:bands_with_surface_potential}(a)]. Interestingly, these end modes are also present in the trivial regime with on-site boundary potentials, as we show in Fig.~\ref{fig:bands_with_surface_potential}(b). Hence, the presence of these additional states cannot be connected to the topology of the bulk.

Let us instead look at the boundary charges of the system which are a protected feature of topological mirror insulators. The \emph{end charge} can be defined as 
the net deviation of the local charge density close to the end from the average
charge density in the bulk~\cite{PYK16}, 
\begin{eqnarray}
Q_\mathrm{L} &=& \lim_{l_0 < L\rightarrow\infty}\sum_{j}^{L} \Theta(l_0-j)(\rho_j-\bar{\rho}), \\
Q_\mathrm{R} &=& \lim_{l_0 < L\rightarrow\infty}\sum_{j}^{L} \Theta(l_0-L+j)(\rho_j-\bar{\rho}),
\end{eqnarray}
where $Q_\mathrm{L}$ and $Q_\mathrm{R}$ denote the left and right end charge, respectively. Here, $L$ is
the length of the chain, $\Theta(x)$ is the Heaviside function and $l_0$ is a cut-off. Moreover,
$\rho_j=\sum_\nu^N|\psi_\nu(j)|^2$ is the local charge density of the
ground state in units of $-e$.
The sum is over all states $\psi_\nu$ up to the chemical potential $\mu$. The quantity $\bar{\rho}$ is the average charge density inside the bulk which is determined by the chemical potential. In the example discussed above, the chemical potential is in the half-filling gap, which amounts to a  bulk charge density of $\bar{\rho}=1$.

Let us now look into the local charge densities of the trivial and topological AAH chains with boundary potentials, which are displayed in the insets of Fig.~\ref{fig:bands_with_surface_potential}. Furthermore, we choose the chemical potential $\mu$ to be above the ``trivial'' end states in the half-filling bulk energy gap. In the topologically trivial regime, the behavior of the local charge density close to the boundary is the same as in the bulk. For the topological mirror insulator, however, the local charge density strongly deviates at the end of the chain and drops to $0$. This already indicates the presence of electronic end charges.
More specifically, the calculated boundary charges are $Q_\mathrm{L}=Q_\mathrm{R}=+1$ in the topological phase
[see Fig.~\ref{fig:bands_with_surface_potential}(a)] and $Q_\mathrm{L}=Q_\mathrm{R}=0$ in the trivial phase [see Fig.~\ref{fig:bands_with_surface_potential}(b)], which is in perfect agreement with the the general relation of Eq.~\eqref{eq:boundary_charge}.

Finally, it can be shown that both the quantized end charges and the end states are stable features of the system even in the presence of nonmagnetic disorder with zero mean~\cite{LBO16}. In other words, as long as the protecting crystalline symmetry is preserved on average, which is reflection for the systems considered here, the topological features remain. However, more general lattice disorder is detrimental to the topological properties of the system. In particular, the quantity that was previously a topological invariant can now assume any rational value and, consequently, the end charges lose their sharp quantization. This is in stark contrast to conventional ``strong'' topological phases whose topology is stable towards any kind of weak lattice disorder.

\section{Conclusions}

The realization that crystalline symmetries have a profound impact on a system's topology has paved the way to a whole new zoo of topological states of matter.
These states of matter fall beyond the standard Altland-Zirnbauer classification since the protecting symmetry is a local spatial symmetry. This symmetry protection, in turn, typically entails the presence of gapless boundary states only on those surfaces which respect the protecting spatial symmetry. 

Here, we have reviewed three different examples of topological insulators protected by crystalline symmetries.
In particular, we have discussed a topological crystalline phase protected by a fourfold rotational symmetry, which features quadratic surface states and is characterized by two $\Z_2$ topological invariants. In crystals with reflection symmetry, on the other hand, it is a nonzero mirror Chern number that gives rise to novel topological insulators whose hallmark is the presence of unpinned surface Dirac cones. Finally, we have also discussed a topological mirror insulator in one dimension where the non-trivial topological properties are reflected in the appearance of quantized end charges, which are stable against weak disorder preserving on average the protecting spatial symmetry.

C.O. acknowledges support from a VIDI grant (Project 680-47-543) financed by the Netherlands Organization for Scientific Research (NWO). A.L. acknowledges support from the Netherlands Organisation for Scientific Research (NWO/OCW), as part of the Frontiers of Nanoscience program.

%\bibliography{references.bib}

%

\end{document}